\documentclass[%
 reprint,
 amsmath,amssymb,
 aps,
prb,
]{revtex4-2}

\usepackage{graphicx}
\usepackage{dcolumn}
\usepackage{bm}

\def\DDDD{$D_{3d}$ }

\def\m{\textsuperscript{--}}

\def\ket#1{\left|{#1}\right\rangle}
\def\bra#1{\left\langle{#1}\right|}

\def\ud{\uparrow/\downarrow}

\def\lambdaeff{\lambda_\mathrm{eff}}
\def\omegaq{\omega_\mathrm{Q}}
\def\omegab{\omega_\mathrm{B}}
\def\toneb{T_1^\mathrm{B}}
\def\toneq{T_1^\mathrm{Q}}
\def\ttwoq{T_2^\mathrm{Q}}
\def\ttwoeff{T_\mathrm{2,eff}^\mathrm{Q}}
\def\tsb{T_\mathrm{S}^\mathrm{B}}

\begin{document}

\preprint{APS/123-QED}

\title{Coherence of Group-IV Color Centers}

\author{Isaac B. W. Harris} 
\email{ibwharri@mit.edu}
\affiliation{Department of Electrical Engineering and Computer Science, Massachusetts Institute of Technology, Cambridge, Massachusetts 02139, USA}

\author{Dirk Englund} 
\email{englund@mit.edu}
\affiliation{Department of Electrical Engineering and Computer Science, Massachusetts Institute of Technology, Cambridge, Massachusetts 02139, USA}

\date{\today}

\begin{abstract}
Group-IV color centers in diamond (SiV\m, GeV\m, SnV\m) have emerged as leading solid-state spin-photon interfaces for quantum information processing applications. However, these qubits require cryogenic temperatures to achieve high fidelity operation due to interactions with the thermal phonon bath.
In this work, we:
(i) derive a detailed model of the decoherence from first-order acoustic phonon processes acting on the spin-orbit fine structure of these color centers;
(ii) demonstrate agreement of the model's predicted coherence times with previous measurements;
(iii) identify regimes to suppress phonon-mediated decoherence by changing magnetic-field and strain bias to allow higher temperature operation.
This methodology enables prediction of decoherence processes in other color centers and solid-state qubit systems coupled to a thermal bath via a parasitic two-level system.
By experiment-anchored decoherence models, we facilitate optimizing qubit coherence for specific applications and devices.

\end{abstract}

\maketitle


\section{\label{sec:intro}Introduction}

Solid-state qubits are promising platforms for quantum computing~\cite{Arute2019,Philips2022,Xue2022} and communication~\cite{Bhaskar2020, Pompili2021, Stas2022} protocols, which require highly coherent quantum information.
However, these qubits couple strongly to the solid-state thermal environment, causing decoherence which is a complicated function of the qubit state, qubit-environment coupling, and the thermal excitations' properties.
Decoherence is typically minimized by cooling to millikelvin temperatures to depopulate thermal excitations, but cryogenic operation limits the cooling power available to remove heat produced during qubit operation.
Since this limitation becomes more severe at colder operating temperature, it is important to understand regimes where qubits are resilient to higher temperature operation.
Detailed modeling of qubit decoherence to predict and understand these regimes is therefore an important need for quantum technologies.

\begin{figure}[ht]
    \centering
    \includegraphics{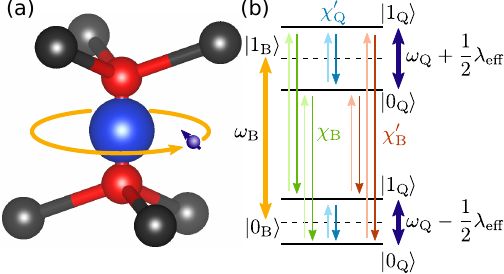}
    \caption{
    (a) Diagram of a group-IV color center composed of a group-IV dopant (blue) at a split-vacancy position in the diamond lattice (carbon atoms shown in black, vacancies in red).
    (b) Energy diagram for the resulting four-level system, with labeled transitions and transition cross-sections corresponding to the branch flipping (green, $\chi_\mathrm{B}$), qubit flipping (blue, $\chi_\mathrm{Q}'$) and qubit+branch flipping (red, $\chi_\mathrm{B}'$).
    }
    \label{fig:intro}
\end{figure}

In this paper, we develop a first-principles decoherence model for one of the leading systems for solid-state quantum spin-photon interfaces, the negatively charged group-IV color centers in diamond~\cite{Hepp2014a, Pingault2017, Trusheim2018, Rugar2020, ArjonaMartinez2022}.
These point defects consist of a group-IV dopant sitting in a split-vacancy configuration between two missing carbon atoms in the diamond lattice (see Fig.~\ref{fig:intro}a).
The color center traps a single charge carrier in localized orbitals, giving it a spin and orbital degree of freedom in the ground state suitable for holding quantum information.
Previous work modeling color center coherence has focused on second-order optical phonon processes~\cite{Norambuena2018, Cambria2023}, or the effect of the nuclear spin bath~\cite{Onizhuk2021a}.
This first-principles modeling of coherence does not capture the main decoherence process for group-IV negatives, which is dominated by first-order absorption and emission of acoustic phonons by the parasitic orbital degree of freedom~\cite{Jahnke2015}.
Phenomenological modeling of the first-order processes with fitted free parameters has also been published~\cite{Jahnke2015, Dhara2023}, and is the basis for current understanding of these color centers' coherence.
However, the lack of detailed first-principles understanding of coherence means that these models are not guaranteed to give accurate predictions when extended to arbitrary bias or environmental conditions.
Wile previous work has shown increases in coherence times by changing strain bias~\cite{Sohn2018} and engineering the phonon density of states~\cite{Chia2021, Klotz2022}, it has not been possible to predict how $T_{1/2}$ coherence times change under arbitrary temperature, strain, magnetic field, and phonon density of states.

The model presented in this paper solves these problems by deriving the first-order phonon-mediated decoherence process from first principles, presented in three parts.
We first derive the model using the group-IV color centers' strain susceptibilities in Section~\ref{sec:theory}, and then compare the predicted coherence times with existing literature in Section~\ref{sec:exp}.
Finally, we discuss optimal bias conditions and associated trade-offs in Section~\ref{sec:bias}, showing that there are regimes with improved coherence not predicted by previous phenomenological models.

\section{\label{sec:theory}Modeling Decoherence Processes}

In the standard spin-orbit basis spanned by the orbit/spin states $\ket{e_\mathrm{g\pm}}_\mathrm{L}\otimes\ket{\ud}_\mathrm{S}$, the group-IV color centers can be modeled with the Hamiltonian~\cite{Hepp2014a}
\begin{equation}\label{eq:spin_hamiltonian}
    \hat{H}_\mathrm{S} = \hat{H}_{\mathrm{SOC}} + \hat{H}_\mathrm{Egx} + \hat{H}_\mathrm{Egy} + \hat{H}_\mathrm{B}  + \hat{H}_\mathrm{L}
\end{equation}
where $\hat{H}_{\mathrm{SOC}}=\lambda_\mathrm{SOC}\sigma_z^\mathrm{L}\sigma_z^\mathrm{S}/2$ is the spin-orbit coupling, $\hat{H}_\mathrm{Egx/y}=-\alpha_\mathrm{Egx/y}\sigma_{x/y}^\mathrm{L}$ is the effect of strain, $\hat{H}_\mathrm{B}=g\mu_\mathrm{B} \boldsymbol{B}\cdot\hat{\mathbf{S}}$ is the spin Zeeman term, $\hat{H}_\mathrm{L} = q\mu_\mathrm{B} B_z \sigma_z^\mathrm{L}$ is the orbital Zeeman term,
$g$ is the electron g-factor, $q$ is the orbital quenching factor, $\mu_\mathrm{B}$ is the Bohr magneton, $\mathbf{B}$ is the applied magnetic field, \mbox{$\hat{\mathbf{S}}=(\sigma_x^\mathrm{S}, \sigma_y^\mathrm{S}, \sigma_z^\mathrm{S})/2$} is the standard electron spin operator, and $\sigma_i^\mathrm{L/S}$ are the Pauli matrices applied to the orbital/spin degree of freedom.
The forms of the four-level Hamiltonian for these perturbations are inferred from group theory~\cite{Tinkham1992, Doherty2011, Hepp2014a}, generally up to a constant factor that must be calculated from first principles~\cite{Thiering2018a} or measured experimentally~\cite{Hepp2014a, Meesala2018a}.

To model decoherence processes, it is convenient to describe the group-IV fine structure as a pair of coupled two-level systems: a qubit degree of freedom and a parasitic branch degree of freedom.
As shown schematically in Fig.~\ref{fig:intro}b, the state of the branch degree of freedom denotes whether the system is in the upper or lower branch, whereas the qubit state dictates whether the system is in the upper or lower level within the branch.
In the system's eigenbasis $\ket{n}, n\in\left\{0,1,2,3\right\}$, we can therefore define $\ket{0_\mathrm{B}0_\mathrm{Q}} = \ket{0}$, $\ket{0_\mathrm{B}1_\mathrm{Q}} = \ket{1}$, $\ket{1_\mathrm{B}0_\mathrm{Q}} = \ket{2}$,
$\ket{1_\mathrm{B}1_\mathrm{Q}} = \ket{3}$.
In this new basis, the four-level system can be parameterized without loss of generality as
\begin{equation}\label{eq:qubit_hamiltonian}
    \hat{H}_\mathrm{S} = \frac{1}{2}\omega_\mathrm{Q}\sigma_z^\mathrm{Q} + \frac{1}{2}\omega_\mathrm{B}\sigma_z^\mathrm{B} + \frac{1}{4}\lambdaeff\sigma_z^\mathrm{Q}\sigma_z^\mathrm{B}
\end{equation}
where $\omega_\mathrm{B}$ is the branch splitting determined by the strength of the spin-orbit coupling and the applied strain, and $\omega_\mathrm{Q}$ is the qubit frequency determined by the applied magnetic field and the effective $g$-factor.
The effective coupling strength $\lambdaeff$ can be viewed as a difference in the qubit frequency $\omega_\mathrm{Q}$ conditioned on the state of the branch degree of freedom.
This coupling originates from the orbital magnetic response $\hat H_\mathrm{L}$, and will go to zero as $B_\mathrm{z}$ goes to zero, or strain becomes much larger than spin-orbit coupling as demonstrated later in Section~\ref{sec:bias}.
For typical group-IV color centers, we can assume that $\omega_\mathrm{B}\gg\omega_\mathrm{Q}\gg\lambdaeff$.

\subsection{Spin-Phonon Coupling}
We first develop a model of the phonon-mediated decoherence of the group-IV color center spin-orbit system.
We restrict the derivation to first-order phonon absorption and emission processes, which are expected to be the dominant source of decoherence for these color centers~\cite{Jahnke2015}.
Since these processes will involve phonons with frequencies of at most $\omegab\approx\lambda_\mathrm{SOC}$, given typical values of $\lambda_\mathrm{SOC} =$ 50~GHz -- 3~THz~\cite{Thiering2018a} we can treat the phonons within the long-wavelength linear elastic limit.
In this regime, thermal phonons apply a strain approximated as being constant over the extent of the color center, whose form depends only on the bulk diamond stiffness.
This is substantially different from decoherence in the NV\m, which is dominated by second-order Raman scattering of quasi-localized optical phonons~\cite{Norambuena2018, Cambria2023}.
The classical displacement due to the phonons can be described as a superposition of plane waves
\begin{equation}
    u_\mathrm{i}(\vec{x},t) = 
    \sum_{\vec{k}\mathrm{m}} \left( 
    A_{\vec{k}\mathrm{m}} q_{\mathrm{i}\,\vec{k}\mathrm{m}}e^{i(k_\mathrm{n} x_\mathrm{n} - \omega_{\vec{k}\mathrm{m}}t)} +
    h.c.
    \right)
\end{equation}
where $A_{\vec{k}\mathrm{m}}$ is the amplitude of the mode with wavector $\vec{k}$ in mode $m\in\left\{ 1, 2, 3 \right\}$.
The $q_{\mathrm{i}\,\vec{k}\mathrm{m}}$ are unit magnitude solutions to the classical phonon eigenmode equation.
\begin{equation}\label{eq:eigenmode}
    \rho\omega_{\vec{k}\mathrm{m}}^2q_\mathrm{i} = c_\mathrm{ijkl}k_\mathrm{j}k_\mathrm{k}q_\mathrm{l}
\end{equation}
where Einstein summation notation is used on the tensor components, $\rho$ is the density of diamond, and $c_\mathrm{ijkl}$ is the Hooke stiffness tensor of diamond.
The modes propagate at a velocity dependent on the direction $\hat k$ equal to $c_{\hat{k}\mathrm{m}}=\omega_{\vec{k}\mathrm{m}} / k$.

To create a quantum-mechanical description of the phonons, we canonically substitute the classical amplitudes for quantum annihilation operators $A_{\vec{k}\mathrm{m}} e^{ - i\omega_{\vec{k}\mathrm{m}}t} \rightarrow \hat a_{\vec{k}\mathrm{m}}$.
Placing the color center at the origin, the strain due to the phonon modes will then be described by the strain operator
\begin{align}\label{eq:strain_operator}
    \begin{split}
        \hat{\epsilon}_\mathrm{uv} =
        \sum_{\vec{k}\mathrm{m}}
        i\sqrt{ 
        \frac{\hbar}{64\pi^3 \rho \omega_{\vec{k}\mathrm{m}}}
        }
        \left(
        k_\mathrm{u} q_{\mathrm{v}\vec{k}\mathrm{m}} + k_\mathrm{v} q_{\mathrm{u}\vec{k}\mathrm{m}}
        \right) \left(
        \hat{a}_{\vec{k}\mathrm{m}} - \hat{a}_{\vec{k}\mathrm{m}}^\dagger
        \right)
    \end{split}
\end{align}

As outlined in equation~\ref{eq:spin_hamiltonian}, the group-IV system is susceptible to strain through the $\hat H_\mathrm{Egx/y}$ terms.
The amount of strain maps to the $\alpha_\mathrm{R}$ strain components as $\alpha_\mathrm{R} = D_\mathrm{Ruv}\epsilon_\mathrm{uv}$, where $R\in\left\{E_\mathrm{gx}, E_\mathrm{gy}\right\}$, and $\mathbf{D}_\mathrm{R}$ are strain susceptibility matrices that depend on two parameters, $d$ and $f$, which are properties of the color center being modeled~\cite{Meesala2018a}
\begin{align}\label{eq:strain_projection}
    \begin{split}
        \mathbf{D}_{E_\mathrm{gx}} =&
        \begin{pmatrix}
        d   &  0 & f/2 \\
        0   & -d & 0 \\
        f/2 &  0 & 0
        \end{pmatrix} \\
        \mathbf{D}_{E_\mathrm{gy}} =&
        \begin{pmatrix}
         0 &  -d  & 0 \\
        -d &   0  & f/2 \\
         0 &  f/2 & 0
        \end{pmatrix} \\
    \end{split}
\end{align}
The total Hamiltonian of the group-IV-phonon system can therefore be modeled as
\begin{equation}\label{eq:H_spin_phonon}
\hat{H}_\mathrm{T} = \hat{H}_\mathrm{S} + \hat{H}_\mathrm{B} + \hat{V}
\end{equation}
The bath Hamiltonian $\hat H_\mathrm{B}$ is the standard phonon field Hamiltonian
\begin{equation}
    \hat{H}_\mathrm{B} = \sum_{\vec{k}\mathrm{m}} \hbar \omega_{\vec{k}\mathrm{m}} \hat{a}_{\vec{k}\mathrm{m}}^\dagger\hat{a}_{\vec{k}\mathrm{m}}
\end{equation}
The coupling $\hat V$ is the coupling between the phonon modes and the group-IV negative system.
Given the phonon-induced strain operator in equation~\ref{eq:strain_operator}, the strain susceptibilities in equation~\ref{eq:strain_projection}, and the effect of the strain on the group-IV system from equation~\ref{eq:spin_hamiltonian}, this coupling reduces to
\begin{equation}\label{eq:spin_phonon_interaction}
    \hat{V} =
    \sum_{\vec{k}\mathrm{mR}} g_{\vec{k}\mathrm{mR}}
    \hat{h}_R
    \hat{P}_{\vec{k}\mathrm{m}}
\end{equation}
where $\hat{h}_\mathrm{Egx/y}=-\sigma_{x/y}^\mathrm{L}$, $\hat{P}_{\vec{k}\mathrm{m}} = i(\hat{a}_{\vec{k}\mathrm{m}} - \hat{a}_{\vec{k}\mathrm{m}}^\dagger)$ and
\begin{equation}\label{eq:coupling_constnt}
    g_{\vec{k}\mathrm{mR}} = \sqrt{ \frac{\hbar}{16\pi^3\rho\omega_{\vec{k}\mathrm{m}}} } D_\mathrm{Ruv} k_u q_\mathrm{v\vec{k}\mathrm{m}}
\end{equation}

The Hamiltonian in equation~\ref{eq:H_spin_phonon} completely describes the group-IV spin-orbit system interacting with a phonon bath to first order.

\subsection{Phonon-Mediated Decoherence}
We now turn to modeling the incoherent evolution of the group-IV color center in the presence of a bath occupied by thermal phonons.
This can be accomplished by assuming that the system and bath states are initially separable, and tracing out the phonon bath under the Born-Markov approximation using standard methods~\cite{Breuer2007}.
The density matrix of the four-level spin-orbit system, $\rho$, then evolves in the interaction picture as
\begin{align} \label{eq:master_unsimplified}
\begin{split}
    \dot{\tilde{\rho}}(t) = 
    \sum_{\vec{k}\mathrm{mR}} \sum_\mathrm{iji'j'} \frac{g_{\vec{k}\mathrm{mR}}^2}{\hbar^2} \left(
       \left[ \hat{s}_\mathrm{Rij} \tilde{\rho}(t), \hat{s}^\dagger_\mathrm{Ri'j'} \right] \Gamma_{\vec{k}\mathrm{m}} e^{i\Delta_\mathrm{iji'j'}t} \right. \\  \left. 
  + \left[ \hat{s}_\mathrm{Ri'j'}, \tilde{\rho}(t) \hat{s}^\dagger_\mathrm{Rij} \right] \Gamma_{\vec{k}\mathrm{m}}^* e^{-i\Delta_\mathrm{iji'j'}t}
    \right)
\end{split}
\end{align}
where $\hat{s}_\mathrm{Rij} = \ket{i}\!\!\bra{i}\hat{h}_\mathrm{R}\ket{j}\!\!\bra{j}$, $\Delta_\mathrm{iji'j'}=(\omega_\mathrm{i} - \omega_\mathrm{j}) - (\omega_\mathrm{i'} - \omega_\mathrm{j'})$, and
\begin{equation}\label{eq:gamma}
    \Gamma_{\vec{k}\mathrm{m}} = \int_0^\infty \mathrm{Tr} \left( \tilde{P}_{\vec{k}\mathrm{m}}^\dagger(t) \tilde{P}_{\vec{k}\mathrm{m}}(t-s) \rho_\mathrm{th}  \right)e^{-i\delta\omega_\mathrm{ij}s} ds
\end{equation}
We neglect for now all terms with $i\neq i'$ and $j\neq j'$ under the rotating wave approximation.
The imaginary part of equation~\ref{eq:gamma} gives a Lamb shift due to the phonon bath that we will assume is already absorbed into other parameters of the system Hamiltonian in equation~\ref{eq:spin_hamiltonian}, and which we can therefore neglect.
Evaluating the real part of the integral in equation~\ref{eq:gamma}, and substituting into equation~\ref{eq:master_unsimplified} results in Lindbladian evolution of the four-level system
\begin{equation} \label{eq:master_nondegenerate_simplified}
    \dot{\tilde\rho}(t) = \sum_\mathrm{ij}\gamma_\mathrm{ij}
    \left( 
       \hat\sigma_\mathrm{ij} \tilde\rho(t) \hat\sigma^\dagger_\mathrm{ij}
       - \frac{1}{2} \left\{ \hat\sigma^\dagger_\mathrm{ij}\hat\sigma_\mathrm{ij},  \tilde\rho(t)\right\} 
    \right)
\end{equation}
with $\hat\sigma_\mathrm{ij} = \ket{i}\!\!\bra{j}$, and
\begin{equation}
    \gamma_\mathrm{ij} =
    2 \pi \sum_\mathrm{R} \left|h_\mathrm{Rij}\right|^2\chi_\mathrm{R}\left| \omega_\mathrm{i}-\omega_\mathrm{j}\right|^3\tilde{n}_\mathrm{th}(\omega_\mathrm{i}-\omega_\mathrm{j})
\end{equation}
and where
\begin{equation}
    \tilde{n}_\mathrm{th}(\omega) = 
    \begin{cases}
        \left(e^{\hbar\omega/k_\mathrm{B}T)} - 1\right)^{-1},& \omega>0 \\
        \left(e^{\hbar\omega/k_\mathrm{B}T)} - 1\right)^{-1} + 1,& \omega<0
    \end{cases}
\end{equation}
is the thermal occupation modified to account for spontaneous emission.
Note that $\chi_\mathrm{R}$, which corresponds to a phonon absorption cross-section, can be calculated using the coupling constants $g_{\vec{k}\mathrm{mR}}$ in equation~\ref{eq:coupling_constnt} from the known strain susceptibilities and diamond phonon modes
\begin{equation}\label{eq:scattering_cross_section}
    \chi_\mathrm{R} = \sum_\mathrm{m}\iint \frac{  \left( D_\mathrm{Ruv} \hat{k}_\mathrm{u} q_{\mathrm{v}\vec{k}\mathrm{m}} \right)^2 }{16\pi^3\rho\hbar c_{\hat{k}\mathrm{m}}^5} d\Omega
\end{equation}
where the integration is over the unit sphere.
Symmetry arguments require that $\chi_\mathrm{Egx} = \chi_\mathrm{Egy} = \chi$.
The first order phonon processes are therefore completely determined by known parameters.

\subsection{Correction for Degenerate Transitions}
In deriving equation~\ref{eq:master_nondegenerate_simplified}, we have assumed that only terms with $i=i'$ and $j=j'$ remain.
However, in the situation where $\lambdaeff$ is very small, there are pairs of transitions where $i\neq i'$ and $j\neq j'$ and $\Delta_\mathrm{iji'j'}\approx0$.
In particular, the qubit flipping and branch flipping transitions highlighted in green and blue in Fig.~\ref{fig:intro}b will have differences in energy of $\lambdaeff$.
In the small $\lambdaeff$ case, additional terms from equation~\ref{eq:master_unsimplified} need to be included.
Converting from the interaction picture to the Schr\"odinger picture, the additional terms are equivalent to performing the substitution
\begin{equation}\label{eq:master_correction_transformation}
    \hat{s}_\mathrm{Rij} \rightarrow \hat{r}_\mathrm{Rij} = \frac{1}{\sqrt{2}}\left( \hat{s}_\mathrm{Rij} + \hat{s}_\mathrm{Ri'j'} \right)
\end{equation}
into equation~\ref{eq:master_nondegenerate_simplified} for each of the degenerate transition pairs $ij/i'j'$.

\begin{figure}[t]
    \centering
    \includegraphics{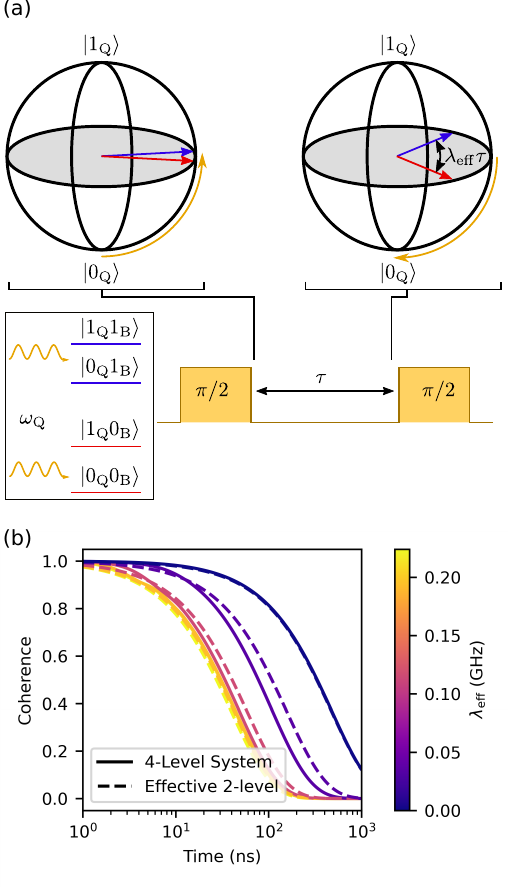}
    \caption{
    (a) Illustration of the evolution of the qubit under a Ramsey sequence in the co-rotating frame conditioned on being in the upper (blue) and lower (red) branch states.
    (b) Evolution of the off-diagonal component of the qubit state as a function of $\lambdaeff$ simulated using the full four-level Lindbladian master equation (solid) and the effective coherence time (dashed) for an SiV\m.
    }
    \label{fig:coherence_correction}
\end{figure}

\subsection{Qubit Coherence}

Equation~\ref{eq:master_nondegenerate_simplified} describes the incoherent evolution of the four-level spin-orbit system.
However, protocols using the group-IV color centers typically ignore the branch degree of freedom, and only use the qubit degree of freedom to store quantum information.
It is therefore convenient to simplify this description to deal with the incoherent evolution of each degree of freedom separately.

We first examine the branch degree of freedom, which we accomplish by treating the qubit as being in a completely mixed state.
Tracing over the product of equation~\ref{eq:master_unsimplified} with $\hat\sigma_\mathrm{z}^\mathrm{B}$ yields
\begin{equation}
    \frac{d}{dt}\langle\hat\sigma_\mathrm{z}^\mathrm{B}\rangle = - 2 \pi \chi_\mathrm{B} \omegab^3 \left( 
    1 + \left( 2n_\mathrm{th}(\omegab) + 1\right) \langle\hat\sigma_\mathrm{z}^\mathrm{B}\rangle \right)
\end{equation}
from which we can infer the characteristic orbital decay rate is then
\begin{equation}
    \frac{1}{\toneb} = 2 \pi \chi_\mathrm{B} \omegab^3\left( 2n_\mathrm{th}(\omegab) + 1\right)
\end{equation}
and an equilibrium value of
\begin{equation}\label{eq:branch_thermal}
    \langle\hat\sigma_z^\mathrm{B}\rangle_\mathrm{th} = -\tanh \left(\frac{\hbar\omegab}{2 k_\mathrm{B}T} \right)
\end{equation}
Here $\chi_\mathrm{B}=\frac{1}{4}\chi\sum_\mathrm{Rij}\left|h_\mathrm{Rij}\right|^2$, is an averaged branch-flipping phonon-scattering cross-section, where the pairs $i,j$ run over the green transitions in Fig.~\ref{fig:intro}b.

Similarly, we can extract coherence times for the qubit degree of freedom by assuming that the branch degree of freedom quickly relaxes to a thermal state determined by equation~\ref{eq:branch_thermal}.
Multiplying equation~\ref{eq:master_unsimplified} by $\hat\sigma_\mathrm{z}^\mathrm{Q}$ and taking the trace gives the phonon-mediated qubit population relaxation time
\begin{equation}
    \frac{1}{\toneq}=\frac{1}{T_1^\mathrm{Q'}} + \frac{1}{T_\mathrm{S}^\mathrm{B'}}
\end{equation}
The $\toneq$ depends on two processes: the first corresponds to direct phonon absorption/emission processes
\begin{equation}
    \frac{1}{T_1^\mathrm{Q'}} = 2\pi\chi_\mathrm{Q}'\omegaq^3\left( 2n_\mathrm{th}(\omegaq) + 1\right)
\end{equation}
and the second qubit-flipping scattering of photons between branches (Orbach process)
\begin{equation}
    \frac{1}{T_\mathrm{S}^\mathrm{B'}} = 4\pi\chi_\mathrm{B}'\omegab^3\left( n_\mathrm{th}(\omegab) +
    \frac{1}{2}\left(1-\langle\hat\sigma_z^\mathrm{B}\rangle_\mathrm{th}\right)\right)
\end{equation}
The $\chi_\mathrm{Q/B}'$ are the corresponding averaged qubit and qubit + branch flipping scattering cross-sections labeled in blue/red in Fig.~\ref{fig:intro}b.

Similarly, we perform the same calculation with $\hat\sigma_\mathrm{x}^\mathrm{Q}$ to find the transverse relaxation time
\begin{equation}\label{eq:T_spin_conserving_scattering}
    \frac{1}{\ttwoq}=\frac{1}{2\toneq} + \frac{1}{2\tsb}
\end{equation}
In addition to being limited by $\toneq$, the transverse decay rate also caries a contribution from the qubit state-conserving phonon scattering
\begin{equation}
    \frac{1}{\tsb} = 4\pi\chi_\mathrm{B}\omegab^3\left( n_\mathrm{th}(\omegab) + \frac{1}{2}\left(1-\langle\hat\sigma_z^\mathrm{B}\rangle_\mathrm{th}\right)\right)
\end{equation}
where $\chi_\mathrm{B}$ is the averaged branch flipping phonon cross-section.
At sufficiently elevated temperatures where the phonons with frequency $\omegab$ are still thermally occupied, $\tsb\approx \toneb$.
We therefore recover the orbital $T_1$-limited qubit $T_2$ previously reported in the literature~\cite{Jahnke2015}.

We have thus far ignored the qubit-branch coupling $\lambdaeff$, effectively assuming it to be very large such that equation~\ref{eq:master_nondegenerate_simplified} is valid.
First, for the case where $\lambdaeff=0$, we repeat the previous calculations for the qubit state with the transformation from equation~\ref{eq:master_correction_transformation}.
We find that $\toneq$ is unaffected; however additional terms from the transformation cancel out the $\tsb$ contribution to $\ttwoq$, leaving $\ttwoq=2\toneq$.
This is to be expected, since if two qubits are completely decoupled and independent from each other, the dynamics of one should not affect the coherence of the other.

We next consider the case where $\lambdaeff$ is non-zero, but sufficiently small that the correction for the near-degenerate transitions cannot be ignored.
\begin{figure*}[ht]
    \centering
    \includegraphics[width=\textwidth]{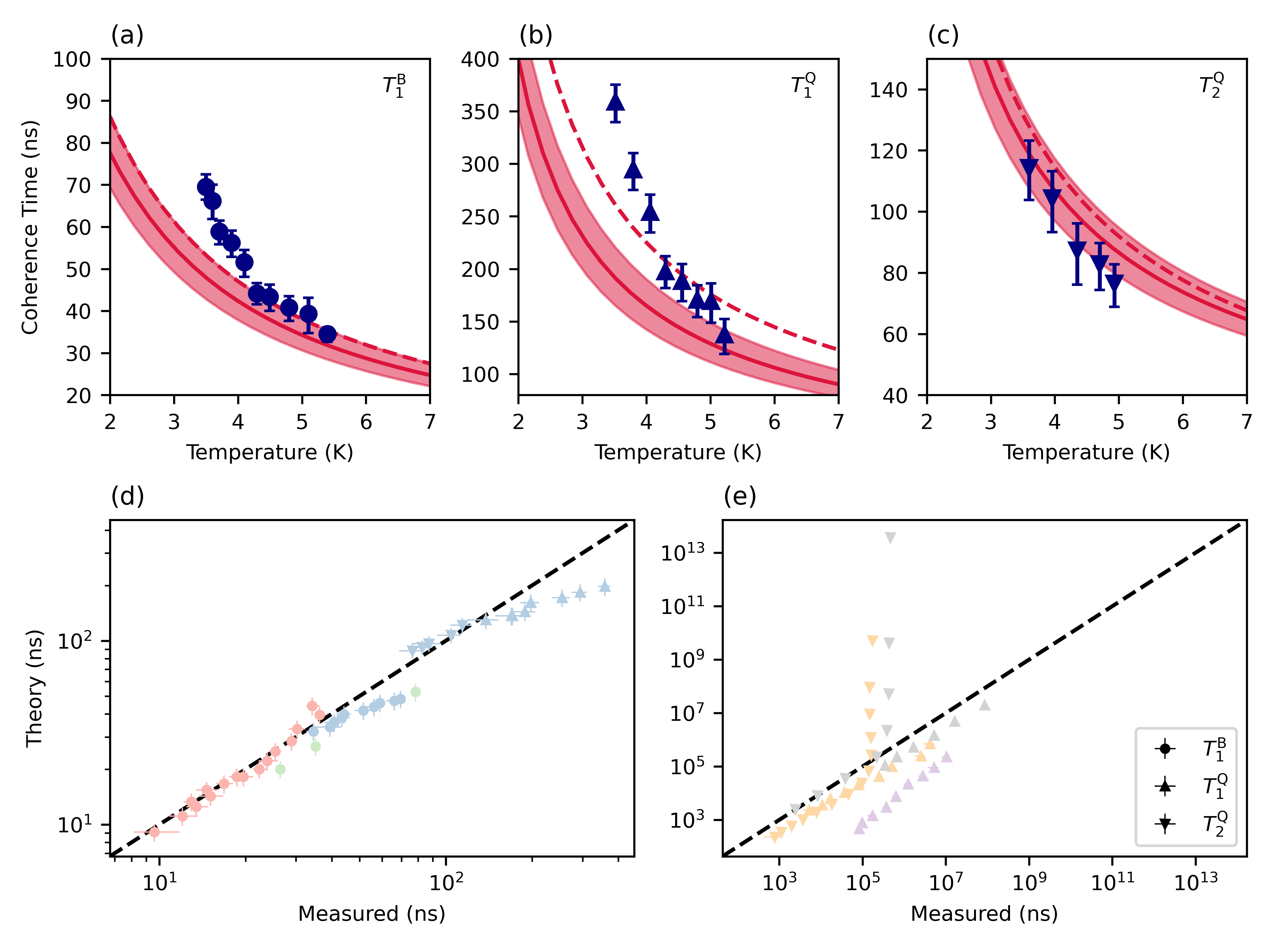}
    \caption{
    (a)-(c) Comparison of theory (red) and experiment (blue) for data for SiV\m\, from Pingault et al.~\cite{Pingault2017} for (a) $\toneb$, (b) $\toneq$, and (c) $\ttwoq$ as a function of temperature.
    Solid lines with fill indicate stated experimental uncertainty in bias conditions and strain susceptibility measurements; dashed line indicates fitted strain susceptibility.
    (d) Plot of measured coherence times vs coherence times predicted for the experimental bias conditions and temperatures for SiV\m from Jahnke et al.~\cite{Jahnke2015} (red), Becker et al.~\cite{Becker2018} (green), Pingault et al.~\cite{Pingault2017} (blue) and for (e) SnV\m from Trusheim et al.~\cite{Trusheim2019} (purple), Rosenthal et al.~\cite{Rosenthal2023} (yellow), Guo et al.~\cite{Guo2023} (grey).
    Circles indicate $\toneb$, upward triangles $\toneq$, and downward triangles $\ttwoq$ for all plots.
    }
    \label{fig:exp}
\end{figure*}
To understand this situation, we first consider the effect of a Ramsey experiment performed on the qubit degree of freedom.
As shown in Fig.~\ref{fig:coherence_correction}a, a microwave pulse brings the qubit state onto the equator of the Bloch sphere.
For a sufficiently strong pulse, we can assume that the qubit state is brought to the same point regardless of the branch state.
The system is then allowed to precess for some time $\tau$, after which the magnitude of the remaining state is measured by applying another microwave pulse to rotate the states back to the poles.
In the frame co-rotating at $\omegaq$, the qubit state will have precessed by an amount $\pm\lambdaeff\tau/2$ depending on if it is in the upper/lower branch.
When the branch state is randomly flipped during the free precession, the degree to which coherence is lost therefore depends on the phase difference between the upper and lower branch, resulting in an oscillating decay rate
\begin{equation}\label{eq:T_spin_corrected}
    \gamma_2^\mathrm{Q}=\frac{1}{2\toneq} + \frac{1}{2\tsb}\left(1-\cos(\lambdaeff\tau)\right)
\end{equation}
In the limit of very large $\lambdaeff\gg0$, $1-\cos(\lambdaeff\tau)$ will oscillate rapidly, averaging out to $\sim1$, while it will go to zero as $\lambdaeff\rightarrow0$.

It is convenient to define an effective coherence time from equation~\ref{eq:T_spin_corrected}.
After some simplifying assumptions shown in Appendix~\ref{sec:effective_coherence_time}, we use the effective coherence time
\begin{equation}\label{eq:T_eff}
    \ttwoeff = 2\toneq\left(\frac{
    1-\exp\left(-\frac{2\pi}{\lambdaeff \toneq}\right)
    }{
    1-\exp\left(-\frac{2\pi}{\lambdaeff} \left(\frac{1}{\toneq}+\frac{1}{\tsb}\right)\right)
    }\right)
\end{equation}
This is illustrated in Fig.~\ref{fig:coherence_correction}b, which shows evolution of $\langle\hat\sigma_\mathrm{x}^\mathrm{Q}\rangle$ for an SiV\m\, under 100~GHz strain, and magnetic applied at various angles to change $\lambdaeff$ with the magnitude varied to keep $\omegaq = 2\pi\cdot 5$~GHz.
The effective two-level exponential decay is shown in dashed lines, alongside the decay from the full four-level system in solid lines.

\section{\label{sec:exp}Comparison to Experiment}

Using the model of first-order phonon-mediated decoherence derived in Section~\ref{sec:theory}, we can now predict the coherence properties of the SiV\m\, and SnV\m\, group-IV color centers.
The relevant parameters for the coherence model are taken from various experimental and theoretical works, and summarized in Table~\ref{tab:known_giv_prop}.
Diamond's relevant elastic properties are also taken from previous work~\cite{Migliori2008}.

\begin{table}[ht]
\caption{Summary of the ground state parameters for SiV\m\, and SnV\m, as well as the resulting phonon scattering cross-sections $\chi$.
Nominal values drawn from \textsuperscript{a}\cite{Goss1996}, \textsuperscript{b}\cite{Hepp2014a}, \textsuperscript{c}\cite{Meesala2018a}, \textsuperscript{d}\cite{Trusheim2018},  \textsuperscript{e}\cite{Guo2023}.}
\label{tab:known_giv_prop}
\begin{center}
\begin{tabular}{c|c|c|c|c|c}
Defect & $\lambda$ (GHz) & q (exp.) & d (PHz) & f (PHz) & $\chi$ ($\times10^{-30}\,s^{-2}$)\\\hline
SiV\m & 50\textsuperscript{a} & 0.1\textsuperscript{b} & 1.3\textsuperscript{c} & -1.7\textsuperscript{c} & 18.1 \\
SnV\m &  830\textsuperscript{d} & 0.15\textsuperscript{d} & 0.787\textsuperscript{e} & -0.562\textsuperscript{e} & 6.2\\
\end{tabular}
\end{center}
\end{table}

With these known parameters, we calculate the phonon scattering cross-section $\chi$, also summarized in Table~\ref{tab:known_giv_prop}, numerically with equation~\ref{eq:scattering_cross_section} using the Lebedev quadrature method~\cite{Burkardt2010,Bast2020}.
We then calculate relevant coherence times at arbitrary temperatures and bias conditions.

A representative comparison between experimentally measured coherence times taken from~\cite{Pingault2017} for SiV\m\, and theoretical coherence times calculated from the independently measured strain susceptibilities are shown for $\toneb$ in Fig.~\ref{fig:exp}a, $\toneq$ in Fig.~\ref{fig:exp}b, and $\ttwoq$ in Fig.~\ref{fig:exp}c as a function of temperature.
We further plot experimental $\toneb$, $\toneq$, and $\ttwoq$ for SiV\m~\cite{Jahnke2015, Pingault2017, Becker2018} in Fig.~\ref{fig:exp}d and for SnV\m~\cite{Trusheim2019, Rosenthal2023, Guo2023} in Fig.~\ref{fig:exp}e against the predicted of coherence time without fitting.
The experimental values are in close agreement with the predictions, on average within 20\% of theory for SiV\m.
If we lift the restriction on using the reported values for magnetic field orientation and strain susceptibility, we can fit the model to the data, reducing the error to less than 10\% (dashed line in Fig. ~\ref{fig:exp}b).
The systematic underestimate of coherence, particularly for $\toneq$, can therefore largely be explained by a combination of a 5$^\circ$ error in magnetic field orientation, in combination with a 5\% error in the strain susceptibility measurement.

The SnV\m\, predictions are also within an order of magnitude of the reported numbers, being underestimated by a factor of $\sim1.7$, suggesting a $\sim30\%$ error in the strain susceptibility parameter, which is only reported from density functional theory calculations~\cite{Guo2023}, and has not been measured experimentally.
The saturation in measured $\ttwoq$ for the predicted $\ttwoq$ greater than $\sim$1~ms in references~\cite{Rosenthal2023, Guo2023} are explained by the presence of nuclear spin bath noise and heating from the pulse sequence, which are not covered by this model.
The source of the underestimate of the $\toneq$ for SnV\m\, in reference~\cite{Trusheim2018} is less clear;
however it may be caused by the change in density of states from the phononic confinement due to the small nanopillars used for optical collection efficiency in this work.

\section{\label{sec:bias}Optimal Bias Conditions}

Having demonstrated the accuracy of the theoretical model, we now turn to predicting the coherence as a function of bias magnetic field and strain.
\begin{figure}
    \centering
    \includegraphics{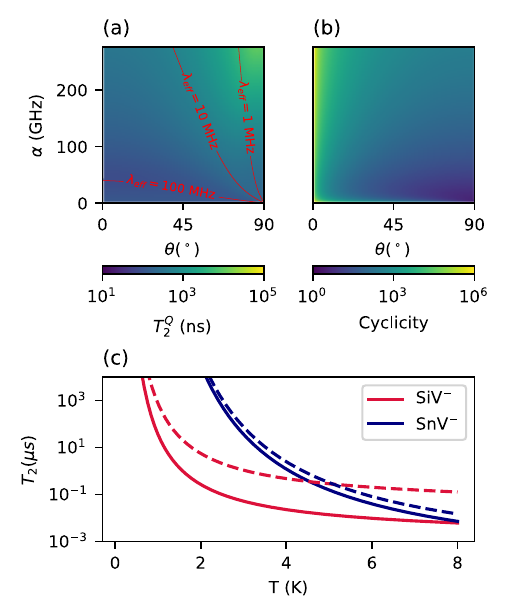}
    \caption{
    (a) Plot of $\ttwoq$ of SiV\m\;as a function of strain and B-field angle from the \DDDD axis at 4~K for a constant 1~GHz qubit frequency.
    Red contours highlight select values of $\lambdaeff$.
    (b) Transition cyclicity for SiV\m\; as a function of strain and $\mathbf{B}$-field orientation.
    (c) Predicted coherence time with 100~GHz of strain with the B-field aligned (solid) and perpendicular (dashed) to the \DDDD~axis as a function of temperature for SiV\m\; and SnV\m.}
    \label{fig:prediction}
\end{figure}
The $\ttwoq$ is shown for an SiV\m\, at 4~K as a function of applied strain and magnetic field angle to the \DDDD axis in Fig.~\ref{fig:prediction}a, where the magnetic field strength is modulated to maintain $\omegaq=2\pi\cdot1$~GHz throughout.
It is clear that $\ttwoq$ is large in the case where the strain ($\alpha$) is large and when the angle of applied magnetic field from the \DDDD axis ($\theta$) approaches 90$^\circ$.
The former is partially due to the decrease in the thermal population of phonons of frequency $\omegab$, which is predicted by previous models~\cite{Jahnke2015} and has been studied experimentally~\cite{Sohn2018, Stas2022}.
However, it is clear from the derivation in Section~\ref{sec:theory} that the decrease in $\lambdaeff$ must also be considered.

As strain increases, the well defined orbital character of the branch states decreases, which suppresses the contribution of the $\hat H_\mathrm{L}$ term.
Similarly, since the $\hat H_\mathrm{L}$ is only dependent on $B_z$, applying a magnetic field off-axis from the \DDDD axis will also decrease its importance in the final energy levels.
Both increased strain and off-axis magnetic field will therefore decrease $\lambdaeff$, as shown in the contours in Fig.~\ref{fig:prediction}a.
In the limits of $\alpha\rightarrow\infty$ or $\theta\rightarrow90^\circ$, the coupling between the branch and qubit degrees of freedom vanish, and $\ttwoq$ is limited by $\toneq$ rather than the much faster phonon-scattering timescale $\tsb$.

The effect of magnetic field orientation on the coherence time suggests an additional parameter that can be optimized to improve color center-based protocols.
This is highlighted in Fig.~\ref{fig:prediction}c, which shows $\ttwoq$ as a function of temperature for SiV\m\, and SnV\m, with 100~GHz of strain.
We show two different magnetic field angle cases:
(1) magnetic field aligned with \DDDD axis (solid line), which is well described by the phenomenological model developed previously~\cite{Jahnke2015}, and 
(2)  magnetic field perpendicular the \DDDD axis (dashed line), which has $\lambdaeff=0$.
A substantial increase from 100s of nanoseconds to a few microseconds is predicted for SiV\m\, in the 1-4~K temperature range at more moderate strains than what has been demonstrated previously~\cite{Sohn2018, Stas2022}.
This increase is not predicted by the phenomenological model, which does not account for the effect of small $\lambdaeff$.
The difference for SnV\m\, is less pronounced, since the larger spin-orbit coupling means that a larger magnetic field must be applied to maintain the 1~GHz qubit frequency, which results in a shorter $\toneq$.
The off-axis magnetic field also causes a decrease in both $\toneq$, as well as optical transition cyclicity, which is shown in Fig.~\ref{fig:prediction}b for SiV\m.
While cyclicities $>10^3$ can be maintained with an off-axis field at moderate (greater than 100~GHz) strain, this is still substantially lower than the large transition cyclicities $>10^5$ that have been demonstrated previously~\cite{Sukachev2017}.
Coherence improvements must therefore be balanced with changes to readout and entanglement fidelity.

\section{\label{sec:conclusion}Conclusion}

The theory outlined in this paper provides a method to quantitatively predict the coherence time of group-IV color centers from independent measurements of their strain susceptibilities.
We have used this theory to: 
\begin{enumerate}
    \item Predict within 20\% the coherence of the SiV\m, and within a factor of $\sim1.7$ the coherence of SnV\m,
    whose strain susceptibilities have been reported in the literature.
    \item Show that there is a bias parameter, the magnetic field orientation, which can have a previously unrecognized effect on the phonon-mediated decoherence process.
\end{enumerate}
Applying an off-axis magnetic field results in an order of magnitude or larger increase in $\ttwoq$, at the cost of decreasing the $\toneq$ and optical transition cyclicity.
This trade-off should be considered in addition to the known trade-offs between initialization speed and $\toneq$ when selecting bias conditions~\cite{Rogers2014}.

This work opens the possibility of optimizing the magnetic angle to balance these trade-offs in a device- or protocol-specific manner~\cite{Dhara2023}.
For example, decreases in optical cyclicity may be mitigated by operating the color center in a cavity with high cooperativity, thus decreasing the probability of the optical readout flipping the spin~\cite{Nguyen2019, Nguyen2019a, Stas2022}.
In addition, a straightforward extension of this work can replace the summation over bulk phonon modes with modes derived from finite element simulations of nanostructures.
This allows for quantitative predictions of the previously studied effects of density of state engineering on group-IV color center coherence~\cite{Chia2021, Klotz2022}.

While we have limited discussion to well-studied emitters, this model is immediately applicable to the other group-IV color centers, the GeV\m~\cite{Inubushi2015} and PbV\m~\cite{Trusheim2019}.
The model is also easily extensible to other novel color centers with similar spin-orbit ground state fine structures such as NV\textsuperscript{0}~\cite{gali:NV:2009, Baier2020, Kurokawa2023} and NiV\m~\cite{Gali2021}, as well as multi-spin hyperfine color centers that couple strongly to an orbital degree of freedom, such as the \textsuperscript{117}SnV\m~\cite{Harris2023, Parker2023a}.
The methods presented here are also applicable to other quantum systems such as color centers in silicon~\cite{Higginbottom2022a}, silicon carbide~\cite{Wolfowicz2019}, rare-earth elements in solids~\cite{Kindem2020}, and other emerging host materials~\cite{Wang2023}, as well as spin qubits in silicon~\cite{Burkard2023} or superconducting qubits~\cite{Blais2021}.
This work therefore derives a general procedure for making quantitative predictions of coherence for systems where a coherent qubit is coupled indirectly to a bosonic environment via a thermalized two-level system.

\begin{acknowledgements}
This work was supported by the STC Center for Integrated Quantum Materials (CIQM) NSF Grant No. DMR-1231319, the National Science Foundation (NSF) Engineering Research Center for Quantum Networks (CQN) awarded under cooperative agreement number 1941583, and the MITRE Moonshot Program.

We would like to thank Prajit Dhara, Kevin C. Chen, Matt Trusheim, and Jesús Arjona Martínez for helpful comments.

\end{acknowledgements}

\appendix

\section{Effective Coherence Time}\label{sec:effective_coherence_time}

We here briefly discuss the derivation of the effective coherence time $\ttwoeff$ in equation~\ref{eq:T_eff} from the oscillating decay rate $\gamma_2^\mathrm{Q}$ in equation~\ref{eq:T_spin_corrected}.

We first note that when the decay rate is a constant $1/\ttwoq$ with an initial value $\langle\sigma_\mathrm{x}^\mathrm{Q}\rangle|_{t=0}=1$, we must have that $\langle\sigma_\mathrm{x}^\mathrm{Q}\rangle = e^{-\tau/\ttwoq}$.
Given that the integral of this quantity from $0$ to $\infty$ is $\ttwoq$, we can define an equivalent effective coherence time for the oscillating decay rate
\begin{equation}\label{eq:T_eff_definition}
    \ttwoeff = \int_0^\infty 
    \left\langle \sigma_\mathrm{x}^\mathrm{Q} \right\rangle dt
\end{equation}

The expectation $\langle\sigma_\mathrm{x}^\mathrm{Q}\rangle$ under the oscillating decay rate evolves as
\begin{equation}
    \frac{d}{dt}\langle\sigma_\mathrm{x}^\mathrm{Q}\rangle = -\gamma_2^\mathrm{Q}(t)\langle\sigma_\mathrm{x}^\mathrm{Q}\rangle
\end{equation}
Under the initial condition $\langle\sigma_\mathrm{x}^\mathrm{Q}\rangle|_{t=0} = 1$, we solve this analytically to yield
\begin{equation}
    \langle\sigma_\mathrm{x}^\mathrm{Q}\rangle = \exp\left(
    -\frac{t}{2\toneq} + \frac{\sin\left( \lambdaeff t\right)}{2\tsb\lambdaeff}
    \right)
\end{equation}
Using the periodic nature of the exponent, we can replace the improper integral used to define the effective coherence time in equation~\ref{eq:T_eff_definition} with
\begin{equation}\label{eq:T_eff_numerical}
    \ttwoeff = \frac{\int_0^{2\pi} 
    \exp\left(
    -\frac{t}{2\toneq} + \frac{\sin\left( \lambdaeff t\right)}{2\tsb\lambdaeff}
    \right) dt}
    {\lambdaeff\left( 1 - \exp\left( -\frac{2\pi}{\lambdaeff}\left( \frac{1}{2\toneq} + \frac{1}{2\tsb} \right) \right) \right)}
\end{equation}
which can be solved numerically.

We can make a further analytical approximation to the integral in the numerator with Laplace's method by expanding the exponent to first order and evaluating to yield equation~\ref{eq:T_eff} in the main text.
This approximation yields the expected coherence time in the limits $\lambdaeff\rightarrow0$ and $\lambdaeff\rightarrow\infty$, however it tends to slightly overestimate the effective coherence time compared to numerical evaluation of equation~\ref{eq:T_eff_numerical} at intermediate values of $\lambdaeff$.
This overestimation is still acceptable, as the oscillating $\gamma_2^\mathrm{Q}(t)$ is small near $t=0$, yielding a larger effective coherence time over short timescales when coherence has not yet been lost.
Since most quantum information protocols are concerned with this highly coherent regime anyway, the analytical approximation to $\ttwoeff$ is still useful.

\bibliography{main}

\end{document}